%% file: main.tex
\documentclass[lettersize,journal,compsoc]{IEEEtran}
\IEEEoverridecommandlockouts
\usepackage{cite}
\usepackage{amsmath,amssymb,amsfonts}
\usepackage{algorithmic}
\usepackage{graphicx}
\usepackage{textcomp}
\usepackage{xcolor}

\def\BibTeX{{\rm B\kern-.05em{\sc i\kern-.025em b}\kern-.08em
    T\kern-.1667em\lower.7ex\hbox{E}\kern-.125emX}}

\usepackage[ruled,linesnumbered]{algorithm2e}

\usepackage{xspace}
\usepackage{multirow}
\usepackage{multicol}
\usepackage{booktabs}
\usepackage{listings}
\usepackage{xcolor}
\usepackage{colortbl}
\usepackage[skins]{tcolorbox}
\tcbuselibrary{skins, breakable, listings, theorems}
\usepackage{subfigure}
\usepackage{bbding}
\usepackage{makecell}

\newcommand{\approach}{RoFTCodeSum\xspace}

\begin{document}

\title{Readability-Robust Code Summarization via Meta Curriculum Learning\\
}

\author{
\IEEEauthorblockN{Wenhao Zeng$^1$, Yitian Chai$^1$, Hao Zhou$^2$, Fandong Meng$^2$, Jie Zhou$^2$, Xiaodong Gu$^1$}\\
\IEEEauthorblockA{
$^1$ \textit{School of Computer Science, Shanghai Jiao Tong University}\\
$^2$ \textit{WeChat AI, Tencent}\\ 
\{zengwh\_cs, xiaodong.gu\}@sjtu.edu.cn}
\thanks{Xiaodong Gu is the corresponding author.}
}


\maketitle

\begin{abstract}

Code summarization has emerged as a fundamental technique in the field of program comprehension. 
While code language models have shown significant advancements, the current models and benchmarks are confined to high-readability code, which contains sufficient semantic cues such as function and variable names. 
In the real world, however, code is often poorly structured or obfuscated, significantly degrading model performance.
In this paper, we first empirically evaluate the robustness of state-of-the-art language models on poor-readability code for the task of code summarization, focusing on (1) their effectiveness, (2) the impact of prompt engineering, and (3) the robustness of different variants.
Experimental results reveal that state-of-the-art models-including GPT-4o and DeepSeek-V3 experience a substantial performance drop when faced with poorly readable code, and that prompt engineering and reasoning-enhanced models offer limited improvements.
Motivated by these findings, we propose \approach, a novel fine-tuning method that enhances the robustness of code summarization against poorly readable code. 
\approach marries the concepts of curriculum learning and meta-learning: based on the original dataset for fine-tuning, it creates curricular training sets, e.g., obfuscating function names and identifiers from the code, respectively, that have progressive difficulty in code comprehension. In each training step, the approach meta-updates the gradients using these progressively challenging datasets, thereby optimizing both accuracy and readability robustness simultaneously.
Experimental results demonstrate that \approach exhibits increased robustness against semantic perturbation while enhancing performance on the original code.

\end{abstract}

\begin{IEEEkeywords}

code language models, code summarization, code readability, meta-learning, curriculum learning

\end{IEEEkeywords}

\section{Introduction}

\IEEEPARstart{C}{ode} summarization has emerged as an indispensable task in the realm of software development. 
It is commonly framed as a machine translation problem, where source code is transformed into natural language descriptions by fine-tuning large language models (LLMs) such as DeepSeek-Coder~\cite{deepseek-coder} and Qwen2.5-Coder~\cite{hui2024qwen2}.

While LLMs have shown significant advancement in this task, their effectiveness relies heavily on the availability of high-readability code, characterized by well-written comments, meaningful variable names, and well-structured syntax~\cite{song2024code, jain2023llm, wang2024your}. 
In real-world development, however, code with poor readability is prevalent~\cite{rani2021comments,shi2022we, bielik2020adversarial}.
 In this paper, we define \textit{poor-readability code} as source code where essential semantic cues are obscured, misleading, thereby hindering comprehension. This is common in various practical scenarios.
For example, in reverse engineering scenarios, developers are often required to interpret decompiled code that lacks meaningful original identifiers. Similarly, in security contexts, malicious code is frequently obfuscated—through the rearrangement of identifiers and structural elements—with the intent of reducing readability and hindering analysis.
Consequently, language models encounter additional challenges in effectively comprehending and processing code with poor readability~\cite{yan2021towards, hu2024effectively, zhou2025beyond}.

\begin{figure}[t]
        \centering
        \includegraphics[scale=0.47, trim=20 10 50 10 clip]{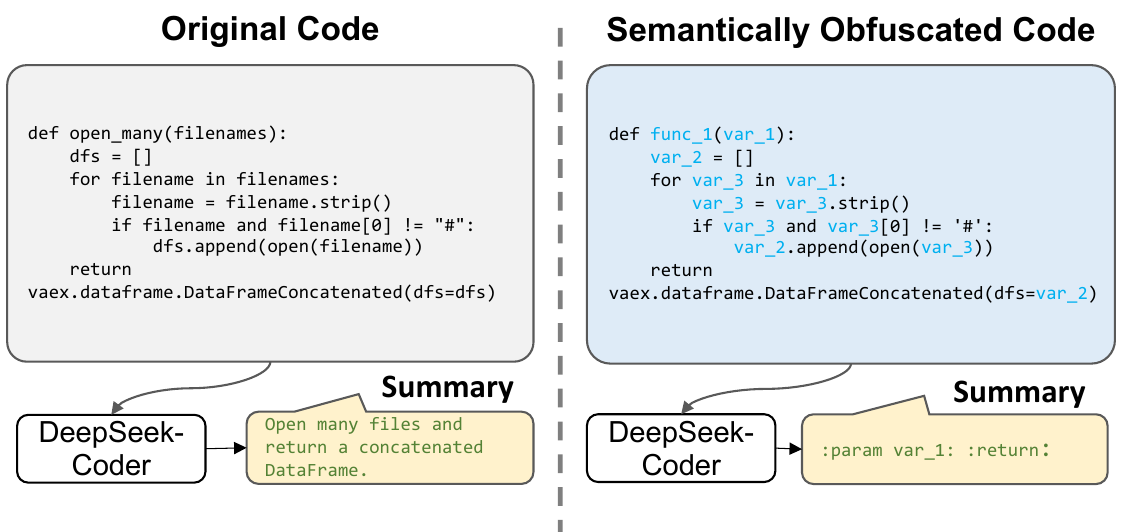}
        \caption{A motivating example of how poor readability interferes with code summarization by language models. The obfuscated identifiers are marked in blue.}
        \label{fig:motivation}
\end{figure}

Figure~\ref{fig:motivation} provides an illustrative example of how readability can impact code summarization. The original code contains meaningful semantic cues such as ``open\_many" and ``filenames", which assist the model in grasping the underlying intent. However, when these variable names are obfuscated, the semantic content of the code remains intact, but the difficulty of comprehension escalates substantially, resulting in inaccurate summaries. 
An ideal code summarization model is anticipated to produce consistent summaries that do not rely on semantic cues. 

To systematically investigate this issue, we conduct an empirical study addressing the following research questions: (1) Are state-of-the-art models effective on poor-readability code? (2) How effective are different prompt engineering methods against poorly readable code? (3) How do different model variants perform on poor-readability code?
We evaluate various LLMs and prompting strategies on both annotated and obfuscated datasets. Our findings reveal substantial performance degradation across all settings, highlighting the urgent need for developing more robust approaches.

Numerous studies have attempted to enhance the robustness of code summarization. For example, Jain et al.~\cite{jain2020contrastive} incorporate contrastive learning into pre-training and enhance the robustness of code representation on code variants. Yan et al.~\cite{yan2021towards} propose a generative tree-based model to enhance the model's ability in understanding the syntax.
However, their approaches rely on creating new training tasks or modifying model architectures, which hinders their compatibility with LLMs, particularly the well-established Transformer architecture and the standard \emph{pretraining-finetuning} pipeline. 

In this paper, we propose \approach (stands for Readability-Robust Fine-Tuning for Code Summarization). \approach aims to improve the robustness of the code summarization model against semantically obfuscated code while enhancing performance on original code.
\approach marries the concepts of curriculum learning and meta-learning: based on the original dataset for fine-tuning, it creates curricular training sets, e.g., obfuscating function names and identifiers from the code, respectively, that have progressive difficulty in code comprehension. In each training step, the algorithm first optimizes the model on the original data with a learning rate $\alpha$. Then, it undertakes meta-training according to the degree of obfuscation from light to heavy (e.g., from FNE to IRN) with learning rate $\beta$ while updating gradients with learning rate $\gamma$.

We conducted experiments using the CodeSearchNet dataset~\cite{husain2019codesearchnet} and employed DeepSeek-Coder~\cite{deepseek-coder} and Qwen2.5-Coder~\cite{hui2024qwen2} as our backbone language models. We compared the performance with baseline methods, including direct fine-tuning, curriculum learning, and CLAWSAT~\cite{yan2021towards}. The results demonstrated that \approach exhibits robustness to semantically obfuscated code while enhancing performance on the original unobfuscated dataset. These findings provide new insights into how to improve code language models in poorly readable code.

The primary contributions of this paper are summarized as follows:

\begin{itemize}
    \setlength\itemsep{0em}
    \item We conduct an empirical study to evaluate the robustness of state-of-the-art LLMs under poor-readability code settings.
    
    \item We proposed \approach, a novel meta-curriculum learning method for enhancing LLMs' robustness to poorly readable code.
    
    \item We extensively experimented with various training methods to demonstrate the efficacy of our approach in code summarization. Our experiments offer valuable insights and practical knowledge for fine-tuning LLMs towards AI-coding tasks.
\end{itemize}

\section{Background}

\subsection{Code Summarization with Language Models}

Code summarization refers to generating natural language comments for a given code snippet~\cite{iyer-etal-2016-summarizing}. It is typically formulated as a machine translation task.
Given the input code $x$ = [$x_1$,..., $x_M$] with $M$ tokens, a large language model such as DeepSeek-Coder~\cite{deepseek-coder} is employed (either by prompt engineering or fine-tuning)
to generate the corresponding comment $y$ = [$y_1$,..., $y_T$].
Given a dataset $D$ = \{($x^i$, $y^i$)\}$_{i=1}^N$ with $N$ samples, the model is usually fine-tuned to minimize the cross-entropy loss~\cite{mao2023cross}:
\begin{equation}
\label{eq:ce}
L(\theta|D) = -\frac{1}{N}\sum_{i=1}^N \frac{1}{T_i} \sum_{t=1}^{T_i} \log p_\theta(y_t^i \mid y_{<t}^i, x^i),
\end{equation}
where $\theta$ denotes the model parameters.

\subsection{Curriculum Learning}

Curriculum Learning (CL) is a training strategy proposed by Bengio et al.~\cite{bengio2009curriculum}, which imitates the natural learning process of humans. It advocates for training a model on progressively increasing levels of difficulty, starting with easy samples and gradually moving towards more complex ones.
The effectiveness of curriculum learning can be understood from two perspectives. Firstly, from an optimization viewpoint, CL can be viewed as a method that firstly optimizes the problem of comparing smoothness, and then gradually optimizes the problem of insufficient smoothness~\cite{bengio2014evolving}.
Secondly, considering the distribution of data, there can often be a bias between the training distribution and the test distribution due to the presence of noisy or mislabeled training data~\cite{gong2016curriculum}. Intuitively, the training and target (test) distributions share a common region with a high density of labeled data and high confidence, which corresponds to the easier samples in CL. By prioritizing the easier samples during training, CL helps align the training and target distributions, reducing the mismatch caused by noisy or mislabeled training data.

\subsection{Meta-learning}

Meta-learning, also known as ``learning to learn," is a subfield of machine learning that focuses on developing algorithms and models that automatically learn how to solve new tasks based on their experience with previous training~\cite{chopra2005learning, snell2017prototypical, finn2017model, nichol2018first}. A meta-learning algorithm typically works by learning a set of ``meta-features" from a training set of tasks, which can then be used to adapt to new tasks more efficiently. 

Model-Agnostic Meta-Learning (MAML) is a popular meta-learning method proposed by Finn et al.~\cite{finn2017model}. MAML is a model-agnostic approach, which means that it can be applied to a wide range of machine-learning models. The basic idea of MAML is to learn a set of model parameters that can be quickly adapted to new tasks with only a few examples. This is achieved by training a meta-learner on a set of tasks, with each task consisting of a small number of examples.

\section{Empirical Study}

\subsection{Study Design}
\label{subsec:study}

To investigate the impact of poor code readability on LLM-based code summarization and to establish a solid motivation for our proposed method, we conduct a preliminary study. This study aims to understand the current landscape of model robustness and is guided by the following research questions:
\begin{itemize}
    \setlength\itemsep{0pt}
    \item \textbf{RQ1: Are state-of-the-art models effective on poor-readability code?} 
    \item \textbf{RQ2: How effective are different prompt engineering methods against poor-readability code?} 
    \item \textbf{RQ3: How do different model variants (base, instruct, reasoning) perform on poor-readability code?}
\end{itemize}
These RQs are fundamental for verifying whether current state-of-the-art models, prompt engineering techniques, and various model variants (including reasoning models) can effectively handle code with poor readability.

To answer these questions, we leverage two datasets: a manually labeled readability dataset (MLRC)~\cite{buse2009learning, scalabrino2016improving} and the widely used CodeSearchNet (CSN) dataset~\cite{husain2019codesearchnet}. MLRC contains human-annotated readability scores. We select a subset of code with associated documentation and divide it into two groups based on the median of the annotated readability scores. Code with scores above the median is categorized as high readability, while code with scores at or below the median is categorized as low readability. We then compare the performance of LLMs on both groups to analyze the impact of readability degradation.
For CSN, we sample the first 2,000 examples from the Python subset of the CSN test set for simplicity and reproducibility and simulate poor code readability through three widely used code obfuscation methods:
\begin{itemize}
    \setlength\itemsep{0pt}
    \item \textbf{Dead Code Injection} (DCI) inserts semantically neutral code that maintains functional semantics \cite{ding2021contrastive, chakraborty2022natgen}. The injected segments include: (1) redundant computations (e.g., unused variable assignments and arithmetic operations), (2) non-functional branches and loops (e.g., dead branches with statically determinable unreachability and conditional statements computing unused variables). 
    \item \textbf{Fuction Name Erosion} (FNE)~\cite{fernandes2018structured, david2020neural, he2018debin} rename the function name with semantic-agnostic symbols. Function name has been shown to be the most critical cue for code comprehension~\cite{siegmund2017measuring,siegmund2014understanding}.
    \item \textbf{Identifier Renaming} (IRN)~\cite{yan2021towards, jain2021contrastive, ding2021contrastive} renames the identifiers in the code snippets with semantic-agnostic symbols such as ``var\_1'' and ``var\_2''. To maintain the minimum readability of the code snippet, we retain the names of external libraries and API calls. 
\end{itemize}


We evaluate a diverse set of state-of-the-art LLMs, including Qwen2.5~\cite{qwen2.5}, DeepSeek-V3~\cite{deepseekai2024deepseekv3technicalreport}, Gemini-2.0-Flash~\cite{gemini20}, Claude-3.5-Sonnet~\cite{claude3-5-sonnet}, and GPT-4o~\cite{openai2024gpt4o} and their variants. These models are selected to represent a range of leading proprietary (e.g., GPT-4o, Claude-3.5) and open-source (e.g., Qwen2.5, DeepSeek-V3) model families, ensuring a comprehensive assessment of the current landscape.
We explore various prompt engineering methods, specifically Zero-Shot, Few-Shot, Chain-of-Thought, and Critique prompting, which are published on GitHub\footnote {https://github.com/Zengwh02/RoFTCodeSum}. We present examples of these prompting methods below.

\begin{tcolorbox}[
float,
floatplacement=htbp,
colframe=black,
colback=white,
coltitle=white,
colbacktitle=black, 
title=Zero-Shot Prompting,
boxrule=0.8pt,
width=0.48\textwidth,
size=small,
fonttitle=\mdseries\small, fontupper=\ttfamily\small, rounded corners]
{\scriptsize
\label{prompt4}
\textcolor{blue}{System Prompt:}

You are a programming assistant skilled at understanding code and generating concise documentation.

\textcolor{blue}{User Prompt:}

Please generate a one-line docstring for the following code that briefly describes its functionality. Only return the docstring without any additional text. \\
Please generate the text in the following format, with triple quotes surrounding the content: \\
"""Generated docstring.""" \\
Code: \\
```python \\
\textcolor{red}{\{code\}} \\
``` \\

\textcolor{blue}{LLM Response:} \\
\textcolor{red}{\{response\}}
}
\end{tcolorbox}

\begin{tcolorbox}[
float,
floatplacement=htbp,
colframe=black,
colback=white,
coltitle=white,
colbacktitle=black, 
title=Few-Shot Prompting,
boxrule=0.8pt,
fonttitle=\mdseries\small, fontupper=\ttfamily\footnotesize, rounded corners]
\label{prompt4}
\textcolor{blue}{System Prompt:}

You are a programming assistant skilled at understanding code and generating concise documentation.

\textcolor{blue}{User Prompt:}

Please refer to the following examples. \\
Example: \\
Code: \\
```python \\
\textcolor{red}{\{fewshot\_code\}} \\
``` \\
Docstring: \\ 
"""\textcolor{red}{\{fewshot\_docstring\}}""" \\

... \\

Please generate a one-line docstring for the following code that briefly describes its functionality. Only return the docstring without any additional text. \\
Please generate the text in the following format, with triple quotes surrounding the content: \\
"""Generated docstring.""" \\
Code: \\
```python \\
\textcolor{red}{\{code\}} \\
``` \\

\textcolor{blue}{LLM Response:} \\
\textcolor{red}{\{response\}}

\end{tcolorbox}

\begin{tcolorbox}[
float,
floatplacement=htbp,
colframe=black,
colback=white,
coltitle=white,
colbacktitle=black, 
title=Chain-of-Thought Prompting,
boxrule=0.8pt,
fonttitle=\mdseries\small, fontupper=\ttfamily\footnotesize, rounded corners]
\label{prompt4}
\textcolor{blue}{User Prompt:}

Code: \\
```python \\
\textcolor{red}{\{code\}} \\
``` \\

Question: \\
1. What is the name of the function? \\
2. What are the input parameters that are being accepted by the function? \\
3. What is the expected output or return value of the function? \\
4. Are there any special requirements or constraints for using the function? \\
5. Does the function have any additional dependencies or external requirements? \\
Please answer the above questions. \\

\textcolor{blue}{LLM Response:} \\
\textcolor{red}{\{response\}} \\

\textcolor{blue}{System Prompt:}

You are a programming assistant skilled at understanding code and generating concise documentation.

\textcolor{blue}{User Prompt:}

Let's integrate the above information. Please generate a one-line docstring for the following code that briefly describes its functionality. Only return the docstring without any additional text. \\
Please generate the text in the following format, with triple quotes surrounding the content: \\
"""Generated docstring.""" \\
Code: \\
```python \\
\textcolor{red}{\{code\}} \\
``` \\

\textcolor{blue}{LLM Response:} \\
\textcolor{red}{\{response\}}

\end{tcolorbox}

\begin{tcolorbox}[
float,
floatplacement=htbp,
colframe=black,
colback=white,
coltitle=white,
colbacktitle=black, 
title=Critique Prompting,
boxrule=0.8pt,
fonttitle=\mdseries\small, fontupper=\ttfamily\footnotesize, rounded corners]
\label{prompt4}
\textcolor{blue}{System Prompt:}

You are a programming assistant skilled at understanding code and generating concise documentation.

\textcolor{blue}{User Prompt:}

Please generate a one-line docstring for the following code that briefly describes its functionality. Only return the docstring without any additional text. \\
Please generate the text in the following format, with triple quotes surrounding the content: \\
"""Generated docstring.""" \\
Code: \\
```python \\
\textcolor{red}{\{code\}} \\
``` \\

\textcolor{blue}{LLM Response:} \\
\textcolor{red}{\{response\}} \\ 

\textcolor{blue}{User Prompt:}

Review your previous answer and find problems with your answer. \\

\textcolor{blue}{LLM Response:} \\
\textcolor{red}{\{response\}} \\ 

\textcolor{blue}{User Prompt:}

Based on the problems you found, improve your answer. \\

\textcolor{blue}{LLM Response:} \\
\textcolor{red}{\{response\}} 

\end{tcolorbox}

To ensure consistency and stability in generation results, we set the temperature parameter to 0.0 during all model inferences. 
We employ two metrics for the evaluation:
\begin{itemize}
    \item \textbf{BLEU-4}~\cite{lin2004orange} is a widely used text similarity metric that measures the overlap between the generated summary and the reference summary by computing the precision of n-grams (up to 4-grams). It is effective in evaluating how closely the generated text matches the reference and has been extensively applied in code summarization tasks~\cite{wang-etal-2021-codet5, lu2021codexglue}.

    \item \textbf{SBERT Similarity}~\cite{reimers-2019-sentence-bert} is a semantic similarity metric that evaluates the equivalence between the generated and reference summaries. It generates sentence embeddings using a pre-trained encoder and computes cosine similarity. In our experiments, we adopt the default implementation and configuration as described in the original paper. In our result tables, we refer to this metric as \textbf{SBERT}.
\end{itemize}

\input{tables/empirical_sota}

\subsection{RQ1: Are State-of-the-Art Models Effective on Poor-Readability Code?}

Table~\ref{tab:sota} illustrates that state-of-the-art LLMs suffer noticeable performance degradation when exposed to poor-readability code. Every tested model, including Qwen2.5-Max-0125~\cite{qwen2.5}, DeepSeek-V3-0324~\cite{deepseekai2024deepseekv3technicalreport}, Gemini-2.0-Flash~\cite{gemini20}, Claude-3.5-Sonnet-20241022~\cite{claude3-5-sonnet}, and GPT-4o-2024-11-20~\cite{openai2024gpt4o}, exhibits consistent reductions in both text similarity (BLEU) and semantic similarity (SBERT) metrics across all poor-readability settings compared to performance on the high-readability code.

The general trend of decline is consistent across both datasets. 
On the MLRC dataset, performance declines are evident when comparing results between high-readability and low-readability code. For example, Qwen2.5-Max drops from 9.33 to 6.30 in BLEU (32.48\%) and from 60.96 to 47.12 in SBERT (22.70\%).
In the CSN setting, Obfuscation methods such as DCI and FNE cause modest reductions. However, methods like IRN lead to sharper declines. For instance, under the IRN setting, BLEU drops by 2.70 for Qwen2.5-Max and 1.92 for GPT-4o. Similarly, SBERT drops across all models, with DeepSeek-V3 experiencing a decrease of 9.81 and Gemini-2.0-Flash dropping from 64.24 to 55.20.

Despite their strong general performance, the most advanced LLMs struggle to effectively comprehend poor-readability code, constraining their applicability in real-world software engineering tasks.

\begin{tcolorbox}[enhanced, colback=white, width=0.99\linewidth, boxrule=1pt,
 left=3pt, right=3pt, top=2pt, bottom=2pt]
State-of-the-art LLMs (e.g., GPT-4o, DeepSeek-V3) exhibit significant and consistent performance degradation when processing poorly readable code, demonstrating their lack of robustness to code clarity reduction.
\end{tcolorbox}

\input{tables/empirical_prompt}

\input{tables/empirical_variants}

\subsection{RQ2: How Effective Are Different Prompt Engineering Methods Against Poor-Readability Code?}

To investigate how effective different prompt engineering methods are against poor-readability code, we conduct empirical experiments using DeepSeek-V3-0324 as the target model. Specifically, we evaluate four commonly used prompting methods for poor-readability code summarization:
\begin{itemize}
    \setlength\itemsep{0pt}
    \item \textbf{Zero-Shot.} Zero-shot prompting adapts LLMs to downstream tasks using simple instructions.
    \item \textbf{Few-Shot.} Few-shot prompting~\cite{gao2023makes} supplements the instruction with some examples when adapting LLMs to downstream tasks.
    \item \textbf{Chain-of-Thought.} Chain-of-thought prompting~\cite{wang2023element} adapts LLMs to downstream tasks by encouraging the model to generate intermediate reasoning steps before producing the final answer.
    \item \textbf{Critique.} Critique prompting~\cite{kim2023language} improves the quality of LLMs’ answers by asking LLMs to find errors in the answers and correct them.
\end{itemize}

We evaluate these prompting methods on two datasets: MLRC and CSN. As MLRC is a small dataset without appropriate examples for few-shot prompting, we only evaluate the other three methods on it.
Consistent with prior findings~\cite{sun2024source}, our results show that there is no guarantee that advanced prompting methods outperform simple zero-shot prompting. As shown in Table~\ref{tab:prompt}, few-shot, chain-of-thought, and critique prompting all exhibit varying degrees of performance degradation compared to the zero-shot prompting. 
Moreover, as code readability decreases under different settings, most prompting methods suffer from notable performance drops. 
On the MLRC dataset, Chain-of-thought prompting achieves only 5.06 BLEU and 41.43 SBERT under low-readability code. 
Notably, the Critique prompting underperforms across all test conditions in the CSN setting, scoring the lowest in both BLEU and SBERT. Specifically, its BLEU drops from 9.28 on the original code to just 7.56 on IRN, while SBERT falls sharply from 56.67 to 46.42.

These findings indicate that prompt engineering offers limited effectiveness when dealing with poor-readability code, highlighting the persistent challenge that obfuscated code poses to LLMs.

\begin{tcolorbox}[enhanced, colback=white, width=0.99\linewidth, boxrule=1pt,
 left=3pt, right=3pt, top=2pt, bottom=2pt]
Prompt engineering shows limited effectiveness in handling poorly readable code, indicating its inability to fully address obfuscated code challenges.
\end{tcolorbox}

\subsection{RQ3: How Do Different Model Variants Perform on Poor-Readability Code?}

To assess how different model variants perform on poor-readability code, we evaluate three variants from the Qwen2.5 series: the base model (Qwen2.5-32B),
the instruction-tuned model (Qwen2.5-32B-Instruct), and the reasoning model (QwQ-32B). Results are presented in Table~\ref{tab:variants}.

In general, the base model exhibits the weakest performance across all settings. Its limited ability to understand the summarization task likely contributes to its poor results.
The instruction-tuned model achieves the best overall scores. Its performance suggests alignment with summarization instruction.
However, its effectiveness degrades significantly on low-readability code, indicating limited robustness to poor readability. 
The reasoning model demonstrates the smallest performance variance across different settings. 
It maintains relatively stable scores under all levels of readability degradation, suggesting that reasoning capabilities help~\cite{zeng2025pruning, zeng2026glimprouterefficientcollaborativeinference}.
However, its overall performance remains below that of the instruction-tuned model.

\begin{tcolorbox}[enhanced, colback=white, width=0.99\linewidth, boxrule=1pt,
 left=3pt, right=3pt, top=2pt, bottom=2pt]
Model variants show divergent performance on poor-readability code: (1) The base model performs worst overall, (2) The instruction-tuned version achieves higher average performance but is highly sensitive to obfuscation, while (3) The reasoning model shows greater stability but slightly lower mean performance than the instruction-tuned variant.
\end{tcolorbox}

\begin{figure}[ht]
        \centering
        \includegraphics[trim=0 0 0 30 clip, scale=0.45]{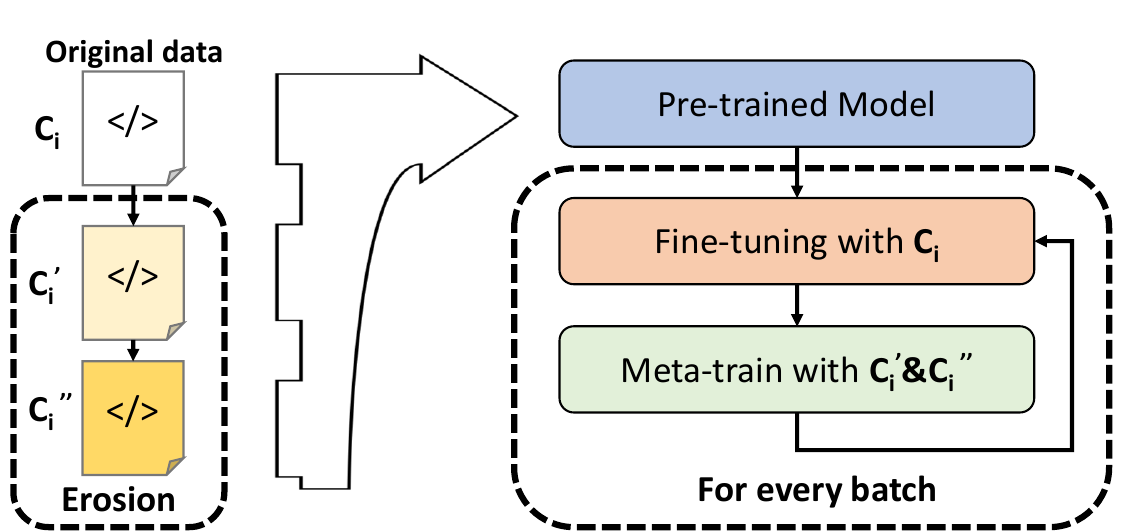}
        \caption{Architecture of \approach, illustrating the flow from the original dataset to progressively more challenging obfuscated datasets.}
        \label{fig:RoFT}
\end{figure}

\begin{figure*}[ht]
        \centering
        \includegraphics[trim=0 0 0 0 clip, scale=0.35]{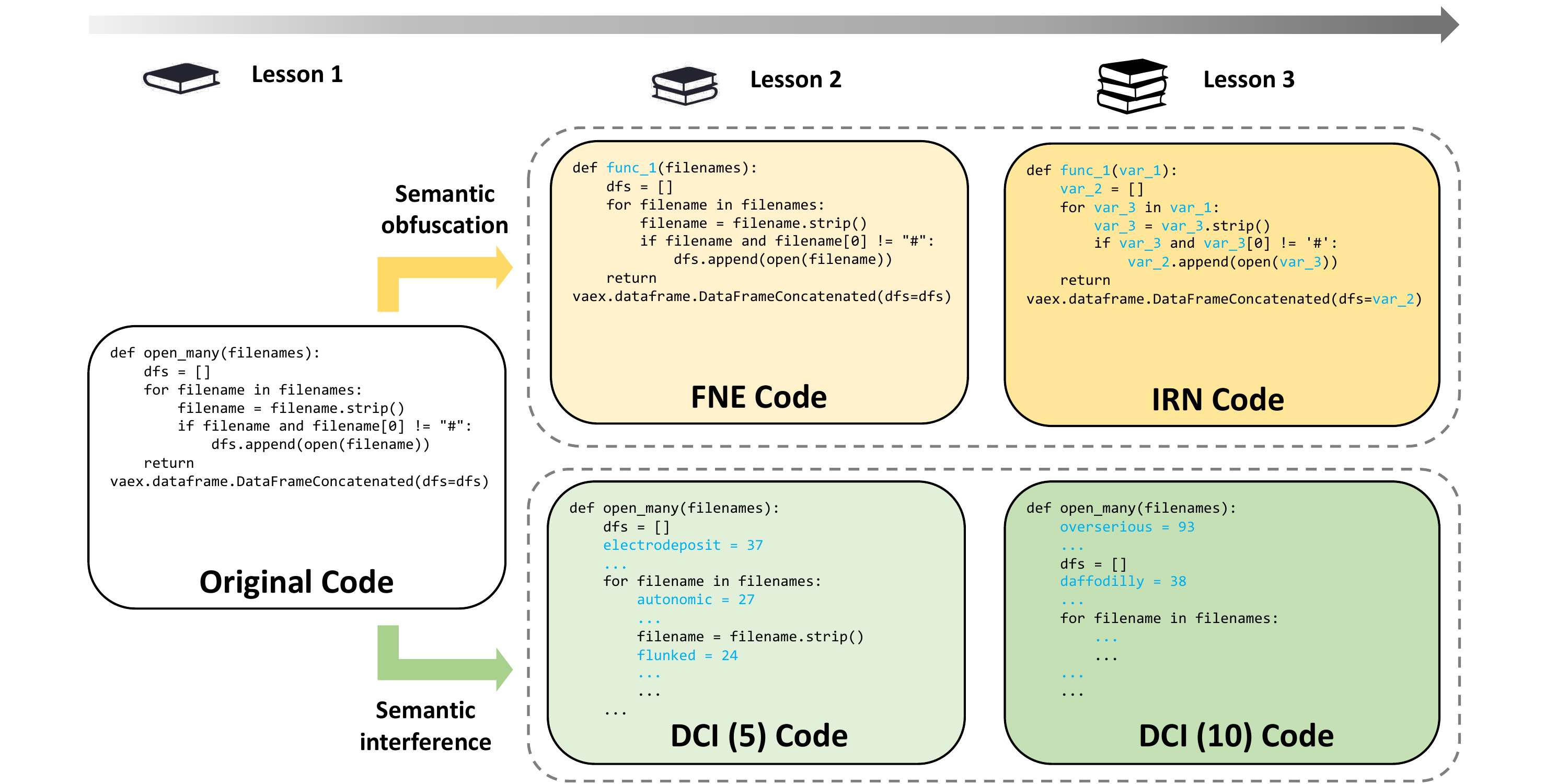}
        \caption{An example of two curricular datasets. The semantically obfuscated parts are marked in blue.}
        \label{fig:erosion}
\end{figure*}

\section{Methodology}

\subsection{Overview}

We regard readability-robust code summarization as an endeavor to enhance the robustness of summarizing poorly readable code while maintaining the performance on high-quality code. Inspired by curriculum learning, our method fine-tunes a language model on multiple summarization tasks. These tasks are carefully constructed to simulate varying levels of code readability, thereby training the model to generalize across a spectrum of real-world code quality. More importantly, the training process of multiple tasks can be equilibrated using meta-learning.

As illustrated in Figure~\ref{fig:RoFT}, the process begins with a standard code summarization dataset (Level 1), which serves as the base for learning on well-structured, readable code. Using the original, readable dataset as the initial level is crucial for anchoring the model's understanding of the primary task, ensuring that robustness is built upon a strong foundation of core summarization capability.
To introduce controlled complexity, we apply code obfuscation techniques such as variable renaming with non-semantic symbols~\cite{yan2021towards, jain2021contrastive, fernandes2018structured, david2020neural} and insertion of dead code~\cite{ding2021contrastive, chakraborty2022natgen}, producing datasets with reduced readability (Levels 2 and 3). We design two distinct curricula based on the type and degree of obfuscation, each comprising three datasets that represent ascending levels of summarization difficulty.

The training process follows a two-stage optimization strategy grounded in meta-learning. Within each training batch, the model first undergoes a standard gradient update using the Level-1 dataset. Then, a meta-update step is performed, leveraging gradients from the more obfuscated Level-2 and Level-3 samples. This joint optimization approach allows the model to retain accuracy on clean code while gaining robustness to obfuscation and other forms of reduced readability.

\subsection{Creating Curricular Datasets}

To facilitate curriculum learning, we construct two sets of datasets with progressively increasing levels of comprehension difficulty. Each curriculum starts from a baseline (original) dataset and incrementally introduces obfuscation to degrade code readability.

We use CodeSearchNet~\cite{husain2019codesearchnet} as the baseline dataset, which is widely adopted in code summarization research~\cite{wang-etal-2021-codet5, feng-etal-2020-codebert, lu2021codexglue}. Our proposed methodology is designed to be language-agnostic, as obfuscation techniques target fundamental code constructs like identifiers and control flow rather than language-specific features. We scope our primary experiments to the Python subset for a focused analysis because it is one of the most popular and widely used programming languages ~\cite{stackoverflow2022survey,tiobeindex,healy2017bridging,hu2026line, shi2024code,shi2024between,li2025swe}.

To simulate increasing difficulty, we reuse the code obfuscation techniques introduced in Section~\ref{subsec:study}—Function Name Erosion (FNE), Identifier Renaming (IRN), and Dead Code Injection (DCI). These techniques are now applied to construct two curricula.

As the most popular way for obfuscating readability, we create our first curricular dataset by degrading semantic cues in the code to increase comprehension difficulty. Starting from the original code, we apply FNE by replacing the function names with semantically uninformative tokens (e.g., \texttt{func\_1}). Next, we apply IRN, which renames all local identifiers (variables, parameters) using generic symbols (e.g., \texttt{var\_1}, \texttt{var\_2}). As previous studies have shown, function and variable names are essential for understanding code intent~\cite{binkley2013impact, yan2021towards, wang-etal-2021-codet5}, and removing these cues forces models to rely more heavily on control flow and logic.  
To ensure a smooth progression, IRN is applied on top of FNE, making it more challenging than FNE alone.

While the semantic-obfuscation curriculum removes helpful information, our second curriculum introduces misleading information. We use DCI to inject dead code into the original functions. By increasing the amount of injected code, we gradually raise the difficulty for understanding the functionality. For instance, DCI~(10) inserts 10 lines of dead code, making it harder than DCI~(5).
All injected code preserves input-output equivalence, ensuring functional correctness while reducing readability.

Figure~\ref{fig:erosion} illustrates an example of curriculum construction by erasing semantic cues. The original code (Left) contains sufficient semantic cues such as ``open\_many'' and ``filenames''. In particular, the function name provides an approximate summary of its purpose, helping language models grasp its general meaning. However, when we obfuscate the function name (Middle), the model has to read through the function body to grasp the main intent. Finally, when we rename all identifiers (Right), the language model needs to analyze the execution of the program before understanding the semantics. The comprehension difficulty increases gradually along the three datasets.

\begin{figure}[t]
        \centering
        \includegraphics[scale=0.40]{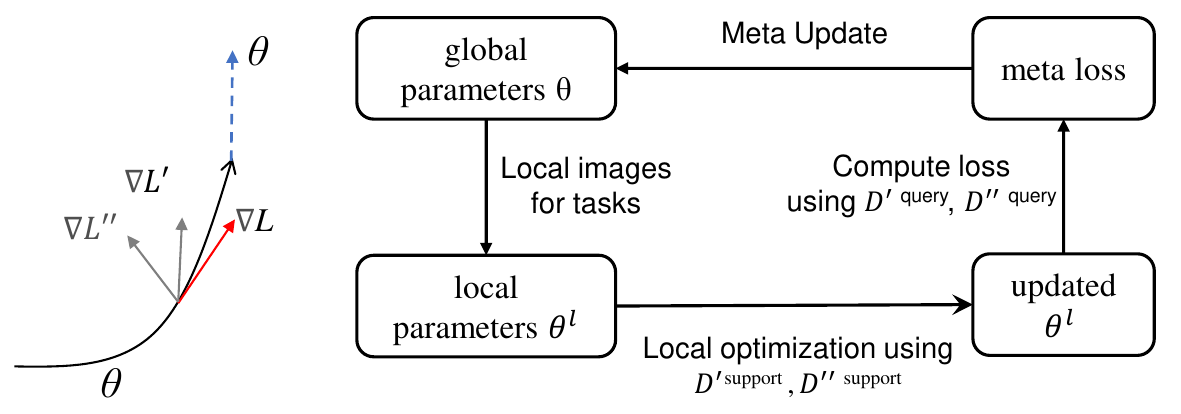}
        \caption{Illustration of the MAML algorithm in our method.}
        \vspace{-10pt}
        \label{fig:maml}
\end{figure}

\subsection{Meta Fine-tuning}

Having created the curricular datasets, we propose a robust fine-tuning process using meta-learning. 
At each training step, we feed the model with batches of data from three sources: the original dataset ($D_i^O$) and two levels of obfuscated datasets with increasing difficulty ($D^{'}_i$ and $D^{''}_i$). 

The robust fine-tuning process consists of two stages.
First, we perform a conventional gradient update on the original dataset $D_i^O$ using the learning rate $\alpha$. This ensures that the model retains a strong grounding in the original task.
Subsequently, we adapt the optimization process through a meta-learning approach to improve robustness against obfuscations. Specifically, we adopt the Model-Agnostic Meta-Learning (MAML) algorithm~\cite{finn2017model}, which aims to learn an optimal initialization that facilitates rapid adaptation to new tasks with a small amount of data. In our setting, the obfuscated code summarization tasks act as source tasks, while the original code summarization task serves as the target task.

For each level of obfuscation (i.e., $D_i^{'}$ and $D_i^{''}$), we create local meta-learning tasks. Each obfuscated batch is split into support and query sets of equal size. We then compute the local gradient $\nabla_{\theta^{l}}L(\theta^{l}|D_i^{query})$ based on the query set after adapting the model using the support set with the learning rate $\beta$. These local gradients are aggregated across tasks to update the global parameters $\theta$ with the learning rate $\gamma$.
The meta-training is performed progressively, starting with lighter obfuscations (i.e., function name removal) and moving to heavier ones (i.e., identifier renaming). Figure~\ref{fig:maml} illustrates the MAML-based training process, and the detailed procedure is presented in Algorithm~1.

\begin{algorithm}[ht]
    \label{algo}
    \small
    \caption{\approach}
    \KwIn{$\alpha$, $\beta$, $\gamma$: step size ; $\theta$: parameters; $D^{O}$: the original dataset; $D^{'}$: level-1 obfuscated data; $D^{''}$: level-2 obfuscated data;}
    \While{not converge}{ 
        Construct three batches \{$D^{O}_{i}$, $D^{'}_{i}$, $D^{''}_{i}$\} that consists of data from $D^{O}$, $D^{'}$, and $D^{''}$, respectively\;
        Run the model on $D^{O}_i$ and calculate the loss $L(\theta|D^O_i)$ using the loss function in Equation~\ref{eq:ce}\;
        Update the global parameters~$\theta$ with gradient descent:\\ \quad\quad$\theta=\theta-\alpha\nabla_{\theta}L(\theta|D^O_i)$\;
        
        \For {$D_{i} \in \{D^{'}_i, D^{''}_i\}$}{ 
            Create a copy of $\theta$ as the local parameters $\theta^{l}$\;
            Split $D_i$ into $\{D_i^\mathrm{support}, D_i^\mathrm{query}\}$\;
             Run a local learning task on $\{D_i^\mathrm{support}, D_i^\mathrm{query}\}$.
             Update local parameters~$\theta^{l}$ with gradient descent:\\ \quad\quad$\theta^{l}=\theta^{l}-\beta\nabla_{\theta^{l}}L(\theta^{l}|D_i^{support})$\;
             Compute the query loss $L(\theta^{l}|D_i^{query})$ \;
            }
        Update the global parameters~$\theta$ using the gradients on the query set:\\ \quad\quad $\theta = \theta-\gamma\nabla_{\theta}\sum_{D^{'}_{i},D^{''}_{i}}L(\theta^{l}|D_{i}^{query})$
        }
\end{algorithm}

\section{Experiment Design}

\subsection{Research Questions}

\textbf{RQ4: How does \approach perform on code summarization?}

\noindent\textbf{RQ5: How do different components of \approach affect the performance?}

\noindent\textbf{RQ6: How do the hyperparameters affect the performance?}

\subsection{Implementation Details}

We conduct experiments on DeepSeek-Coder~\cite{deepseek-coder} and Qwen2.5-Coder~\cite{hui2024qwen2}, two cutting-edge LLMs for code intelligent tasks.
We fine-tune the 1.3B and 1.5B versions, respectively, and reuse checkpoints released by Huggingface \footnote{https://huggingface.co/Qwen/Qwen2.5-Coder-1.5B}\footnote{https://huggingface.co/deepseek-ai/deepseek-coder-1.3b-base} with the default hyperparameter settings. 

The maximum lengths for the source and target sequences are set to 256 and 128, respectively. The model was trained with a batch size of 64 and a weight decay of 0.05 for 3 epochs. For non-meta-learning approaches, we employed a linear learning rate scheduler with an initial rate ($\alpha$) of 5e-5, incorporating 1000 warmup steps to stabilize early training. For meta-learning methods, we set three learning rates ($\alpha$, $\beta$, and $\gamma$) to 5e-5. Following meta-learning conventions that inherently optimize initial parameter sensitivity, we eliminated the warmup phase (0 warmup steps).

We measure the performance on the validation set during training and select the last checkpoint of the model for the baseline of curriculum learning.
For the final evaluation, we select the checkpoint that attains the highest average accuracy on the validation sets of all datasets within each curriculum.

\input{tables/result_fne_irn_dscoder}
\input{tables/result_dci_dscoder}

\input{tables/result_fne_irn_qwen}
\input{tables/result_dci_qwen}

\subsection{Baselines}

We compare \approach with baseline methods used for fine-tuning LLMs and similar methods using curriculum learning.

    \begin{enumerate}
        \setlength\itemsep{0pt}
        \item \textbf{Zero-shot}, directly leverages the base code LLM’s fill-in-the-middle capability for code summarization without any fine-tuning.
        \item \textbf{Fine-tuning}, applies conventional fine-tuning to the LLM using the original code summarization training dataset.
        \item \textbf{Fine-tuning-all}, fine-tunes the LLM on all adversarially obfuscated curricular datasets, effectively enhancing model robustness through exposure to adversarial examples.
        \item \textbf{Curriculum Learning}~\cite{bengio2009curriculum}, sequentially fine-tunes the LLM on datasets of increasing complexity, progressing from the original code dataset to the FNE dataset and finally the IRN dataset.
        \item \textbf{CLAWSAT}~\cite{jia2023clawsat}, implements robust training by alternating between original and adversarially obfuscated curricular datasets within each epoch.
    \end{enumerate}
    
We implement all baselines based on the backbone LLMs (i.e., DeepSeek-Coder and Qwen2.5-Coder) with the default hyperparameters.

\section{Results and Analysis}

\subsection{RQ4: How does \approach perform on code summarization?}

Tables~\ref{tab:result:dscoder_fne_irn_curriculum} and \ref{tab:result:dscoder_dci_curriculum} compare the performance of various methods in the two curricula, respectively. In addition, we present results based on Qwen2.5-Coder in Tables~\ref{tab:result:qwencoder_fne_irn_curriculum} and \ref{tab:result:qwencoder_dci_curriculum}.

The experimental results (Tables~\ref{tab:result:dscoder_fne_irn_curriculum}, \ref{tab:result:dscoder_dci_curriculum}, \ref{tab:result:qwencoder_fne_irn_curriculum} and \ref{tab:result:qwencoder_dci_curriculum}) demonstrate the substantial superiority of \approach over both fine-tuning and curriculum learning baselines across different models. The improvements of \approach are statistically significant across all results (Wilcoxon signed-rank test, $p < 0.05$). We have three key observations:

First, while baseline methods suffer from significant performance degradation on obfuscated datasets, \approach consistently maintains robust performance with minimal decline, indicating enhanced resilience to semantic obfuscation and interference. This trend holds for both DeepSeek-Coder and Qwen2.5-Coder, affirming the generality of the approach.

Second, quantified by the average improvement metric~\cite{zhan2021meta}, \approach achieves a notable +3.31 BLEU improvement under semantic obfuscation with DeepSeek-Coder, representing a 35.7\% enhancement over FT\textsubscript{All} (+2.44). With Qwen2.5-Coder, a similar pattern emerges: \approach outperforms FT\textsubscript{All} by +1.92 vs. +1.38. Furthermore, in semantic-interference settings, \approach achieves nearly twice the improvement of FT\textsubscript{All} on both models (e.g., +1.32 vs. +0.66 with DeepSeek-Coder and +1.47 vs. +0.77 with Qwen2.5-Coder).

Notably, \approach not only preserves but even improves performance on original code, demonstrating its dual advantage. With DeepSeek-Coder, it achieves a BLEU score of 24.94, surpassing FT\textsubscript{All} (24.39) by 0.55. Similarly, on Qwen2.5-Coder, it reaches 23.78, outperforming FT\textsubscript{All} (23.28). These results highlight that \approach enhances robustness against obfuscated code while improving performance on original code.

\begin{tcolorbox}[enhanced, colback=white, width=0.99\linewidth, boxrule=1pt,
 left=3pt, right=3pt, top=2pt, bottom=2pt]
\approach exhibits high performance across both original and poor-readability code, showing strong robustness to semantic obfuscation and interference. 
\end{tcolorbox}

\subsection{RQ5: How do different components of \approach affect the performance?}

Compared with methods relying solely on curriculum learning (CL and CLAWSAT), \approach (DeepSeek-Coder) achieves 1.38 and 1.64 higher average BLEU than CL and CLAWSAT, respectively, in the semantic-obfuscation curriculum. Similarly, in the semantic-interference curriculum, \approach outperforms CL and CLAWSAT by 1.04 and 1.28 BLEU on average.

When compared to fine-tuning methods, \approach exhibits consistent superiority. Against FT\textsubscript{All}, which directly fine-tunes on all data, \approach achieves +0.87 and +0.66 BLEU gains in the semantic-obfuscation and semantic-interference curricula, respectively.

The results on Qwen2.5-Coder are consistent with those observed on DeepSeek-Coder: in the semantic-obfuscation curriculum, \approach surpasses CL and CLAWSAT by 0.93 and 1.17 BLEU, and outperforms FT\textsubscript{All} by +0.54 BLEU. Similarly, under the semantic-interference curriculum, \approach achieves 1.12 and 1.09 BLEU improvements over CL and CLAWSAT, and exceeds FT\textsubscript{All} by +0.70 BLEU.

The results confirm the synergistic effect of combining curriculum learning with meta-learning across different code language models. While curriculum learning (CL/CLAWSAT) incrementally exposes the model to increasingly challenging samples, meta-learning enables rapid adaptation to new obfuscation patterns. Our method leverages meta-training on a curriculum, allowing the model to generalize more effectively to unseen examples by focusing on meta-level features such as structural logic rather than surface-level cues like variable names.

\begin{tcolorbox}[enhanced, colback=white, width=0.99\linewidth, boxrule=0.8pt,
 left=2pt, right=2pt, top=2pt, bottom=2pt, drop fuzzy shadow=black,]
The performance gains of the \approach stem from the effective integration of curriculum learning and meta-learning. 
\end{tcolorbox}

\begin{figure}[!h]
    \centering
    \subfigure{
        \label{fig:fig1}
        \includegraphics[width=0.35\textwidth, trim=100 20 70 10 clip]{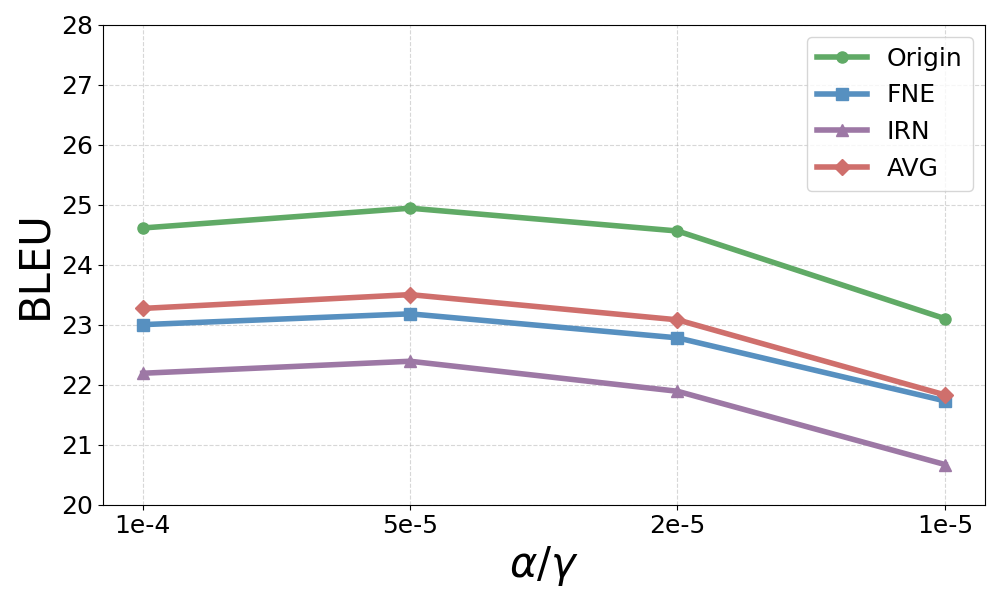}
   }\hfill
    \subfigure{
        \label{fig:fig2}
        \includegraphics[width=0.35\textwidth, trim=100 20 70 10 clip]{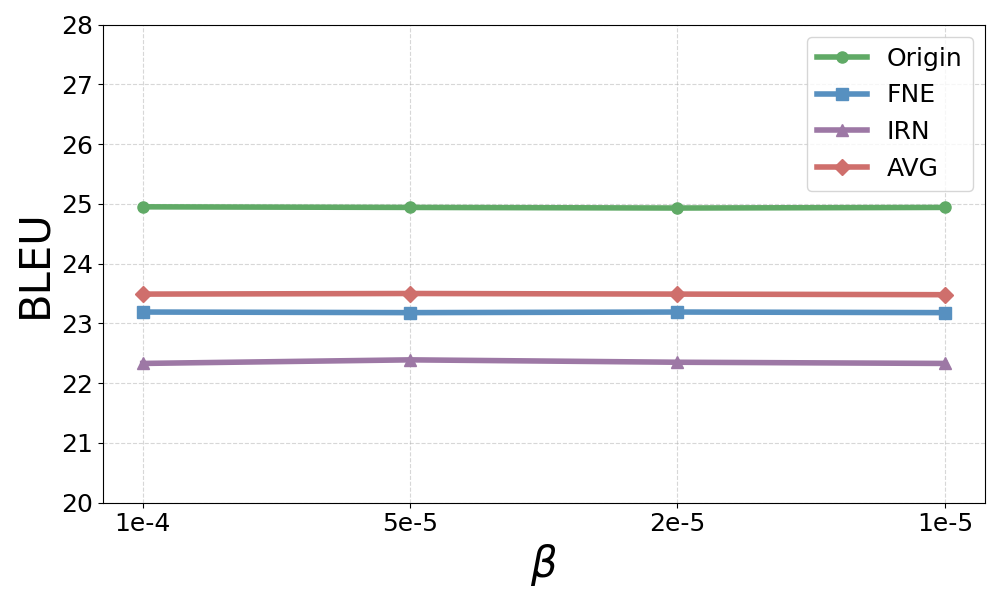}
    }
    \caption{Performance of \approach (DeepSeek-Coder) under different hyperparameters in the semantic-obfuscation curriculum}
    \label{fig:fig0}   
\end{figure}

\begin{figure*}[!h]
    \centering
    \includegraphics[scale=0.28]{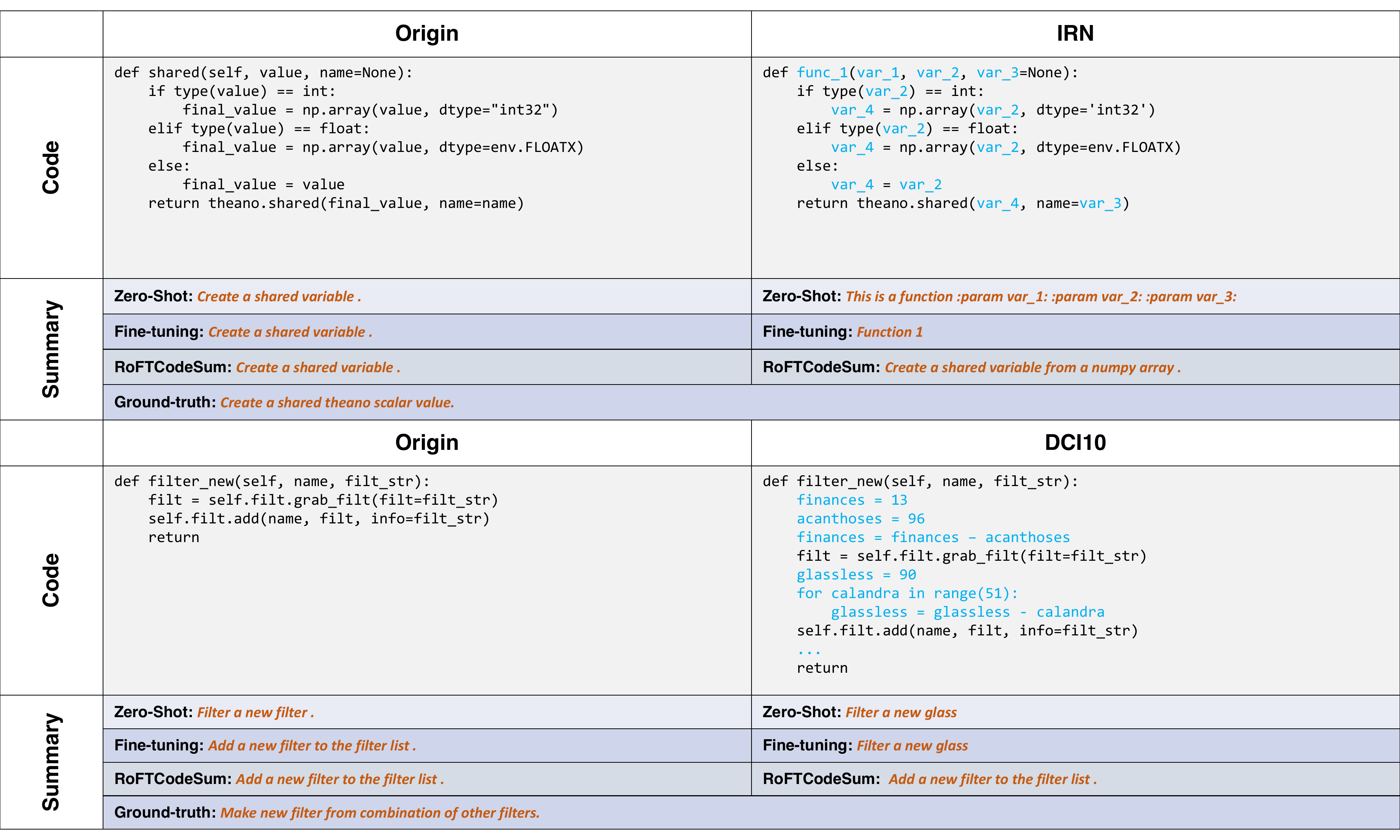}
    \caption{Two cases of readability robust code summarization by \approach (DeepSeek-Coder). The obfuscated tokens are marked in blue.}
    \vspace{-10pt}
    \label{fig:case}
\end{figure*}

\subsection{RQ6: How do the hyperparameters affect the performance?} 

Figure~\ref{fig:fig0} shows the performance of \approach under different learning rates $\alpha$, $\beta$ and $\gamma$ in the semantic-obfuscation curriculum.
To balance the influence across different datasets in the curriculum, we set $\alpha$ and $\gamma$ to identical values. We conduct systematic experiments on the learning rate parameters by evaluating $\alpha$ / $\gamma$ and $\beta$ across {1e-4, 5e-5, 2e-5, 1e-5}. Experimental results demonstrate optimal performance when both $\alpha$ / $\gamma$ and $\beta$ are set to 5e-5. Notably, model outcomes exhibited significant sensitivity to variations in $\alpha$ / $\gamma$ values, while remaining relatively robust to $\beta$ adjustments. This empirical analysis motivates our final parameter selection of $\alpha$ = $\gamma$ = 5e-5 and $\beta$ = 5e-5, which optimally balances convergence stability and task performance across all datasets.

\begin{tcolorbox}[enhanced, colback=white, width=0.99\linewidth, boxrule=1pt,
 left=3pt, right=3pt, top=2pt, bottom=2pt]
\approach achieves optimal performance when $\alpha = \gamma = 5\text{e--}5$ and $\beta = 5\text{e--}5$. Performance is more sensitive to $\alpha/\gamma$ than to $\beta$.
\end{tcolorbox}

\subsection{Case Study}

Figure~\ref{fig:case} shows two cases of code summarization by \approach and direct fine-tuning in the semantic-obfuscation and -interference curricula, respectively. In the first case, both \approach and fine-tuning initially generate identical summaries ``Create a shared variable'' for the code snippet before semantic obfuscation is introduced. However, after applying semantic obfuscation, the fine-tuning approach produces an overly simplified and semantically incomplete summary ``Function 1'', while the \approach maintains better semantic comprehension by generating ``Create a shared variable from a numpy array''. In the second case, both approaches initially produce identical summaries ``Add a new filter to the filter list'', demonstrating comparable baseline performance under normal conditions. However, when semantic interference is introduced, the fine-tuning approach generates ``Filter a new glass'', which shows significant semantic drift and appears to be conflating unrelated concepts. In contrast, \approach maintains its original summary ``Add a new filter to the filter list'', demonstrating remarkable consistency despite the interference. The findings demonstrate the effectiveness of \approach in enhancing the robustness across diverse obfuscated code.

\section{Discussion}

Our experimental results demonstrate the effectiveness of \approach, but they also raise broader questions about its implications and limitations.

\subsection{Principal Findings}
Our primary finding is that \approach not only enhances robustness on poor-readability code but also improves performance on original, high-quality code. This "win-win" outcome challenges the common assumption that robustness training often involves a trade-off with general performance. We hypothesize this is because the curriculum of progressively harder obfuscations forces the model to move beyond superficial semantic cues (like variable names) and learn the underlying control flow and algorithmic logic of the code.

\subsection{Implications for Practitioners and Researchers}

For software engineering practitioners, our work marks a step towards transforming LLM-based coding assistants from tools that work well under ideal conditions into reliable partners for the complexities of real-world software maintenance. The enhanced robustness offered by \approach means that developers can more confidently apply these models to legacy systems, third-party libraries, or any codebase where coding conventions are inconsistent or documentation is sparse. In critical domains like security analysis and reverse engineering, where code is often intentionally obfuscated, a model trained with our method could provide a crucial first-pass analysis, significantly reducing the manual effort and cognitive load required to understand malicious or protected software. Ultimately, this leads to more predictable and trustworthy AI-powered tools that can simplify maintenance and improve code comprehension across a wider spectrum of software projects.

For the research community, \approach provides more than a solution for a specific task; it offers a generalizable paradigm for robust fine-tuning. The core principle of marrying meta-learning with a curriculum of progressively challenging input variations is not limited to code summarization. This framework can be readily adapted to enhance robustness in other critical code intelligence tasks. For example, in defect detection, it could train models to identify logical bugs irrespective of variable naming styles; in code translation, it could help migrate legacy code by focusing on algorithmic equivalence rather than superficial syntax. This opens up several avenues for future research, including investigating optimal curriculum design, exploring the framework's applicability to other forms of code variance (e.g., stylistic differences, noisy data).

\subsection{Generalization to Unseen and Complex Obfuscations}

Our current curricula focus on two common forms of poor readability: semantic obfuscation and structural interference. A critical question is how \approach would generalize to entirely unseen, structurally-complex obfuscations, such as Control-Flow Flattening (CFF).

We hypothesize that our method provides a degree of generalization, but would not be fully robust. For instance, because our IRN curriculum forces the model to trace program logic from control flow rather than relying on semantic cues, it might be more resilient than a baseline model when faced with CFF. It has already learned to ignore misleading names and focus on structure.

However, advanced techniques like CFF introduce specific, highly complex structural patterns, such as transforming all loops and conditionals into a large switch statement inside a single while loop, which the model has never encountered during training. We cannot expect the model to fully understand this drastic transformation without seeing examples. Therefore, to achieve full robustness against such a specific class of obfuscation, it would need to be explicitly integrated into our training. A key advantage of our \approach framework is its extensibility: one could readily introduce CFF-obfuscated code as a new, more difficult level within the curriculum. We leave the exploration of this continual learning for an expanding set of complex obfuscations as a key direction for future work.

\section{Threats to Validity}

\subsection{External Validity}
This aspect concerns the generalizability of our findings. 
\textbf{(1) Language Scope:} Although our main experiments were conducted on Python, our proposed methodology is designed to be language-agnostic. This is because the obfuscation techniques we target primarily manipulate fundamental code constructs, such as identifiers and control flow. Therefore, we expect our approach to generalize to other programming languages (e.g., Java or JavaScript) with minimal adaptation. \\
\textbf{(2) Synthetic vs. Real-world Data:} 
Our study primarily uses synthetically obfuscated code (IRN, DCI). This approach is essential for a controlled and reproducible experiment, allowing us to measure performance against specific types of readability degradation precisely. We mitigated this threat partially by choosing obfuscations that common, high-impact issues found in the wild: Identifier Renaming (IRN) simulates the uninformative variable names prevalent in legacy code, while Dead Code Injection (DCI) simulates the non-functional, distracting logic often left behind during protracted maintenance cycles.
However, we acknowledge that the patterns and complexity of synthetic obfuscation may not fully capture the entire spectrum of naturally occurring low-readability code. To more comprehensively address this threat in the future, a valuable contribution would be the curation of a new, large-scale benchmark of verified real-world low-readability code. This would involve mining open-source repositories to identify code with high complexity, low readability scores, or code from projects known for their "code smells" (e.g., legacy systems). Evaluating \approach and other models on such a benchmark would provide definitive evidence of their practical utility in real-world software maintenance scenarios.

\subsection{Internal Validity}
The primary internal threat is the selection of the key hyperparameters in our \approach framework, including the three learning rates that govern our meta-curriculum optimization: the standard update rate for the original data ($\alpha$), the inner-loop adaptation rate for the obfuscated tasks ($\beta$), and the outer-loop meta-update rate ($\gamma$). The balance between these three rates is critical for simultaneously maintaining accuracy on clean code and building robustness on obfuscated code.
We mitigated this threat by conducting a sensitivity analysis (as presented in RQ6), demonstrating that our chosen configuration is both stable and effective, and that performance is more sensitive to $\alpha/\gamma$ than to $\beta$. While this analysis provides confidence in our results, we acknowledge that our grid search was not exhaustive, and other configurations might exist that could further optimize this performance trade-off.

\subsection{Construct Validity}
This relates to whether our evaluation metrics accurately measure code summarization quality. We employ a combination of text-level and semantic-level metrics, specifically BLEU-4 and SBERT, to capture different dimensions of similarity between generated and reference summaries. While these metrics are widely adopted in the literature, they remain imperfect proxies for human judgment of readability, coherence, and usefulness. A complementary human evaluation would provide a more comprehensive assessment of model performance.

\section{Related Work}

\subsection{Code Summarization}
Code summarization aims to comprehend code and automatically generate concise natural language descriptions for source code snippets, thereby assisting developers in program understanding, maintenance, and documentation.~\cite{iyer-etal-2016-summarizing}

Shi et al.~\cite{shi2022evaluation}  conducted a systematic and in-depth analysis to understand the progress and limitations of neural code summarization. Specifically, they evaluated five state-of-the-art neural models across six BLEU metric variants, four code preprocessing operations (and their combinations), and three widely used datasets.
In the era of large language models (LLMs), Sun et al.~\cite{sun2024source} carried out a comprehensive investigation into the capabilities of LLMs for code summarization. Their work examined the impact of various prompting techniques and model generation settings. Notably, smaller models such as CodeLlama-Instruct (7B) were shown to outperform GPT-4 in certain code summarization tasks.
Despite these advancements, understanding poorly readable or obfuscated code remains a major challenge. Hu et al.~\cite{hu2024effectively} conducted a detailed robustness study focusing on code readability. They constructed seven obfuscated datasets derived from standard benchmarks to systematically evaluate how code summarization models respond to varying levels of readability. Their findings revealed that current models are vulnerable to low-readability code, with performance deteriorating significantly in obfuscated settings. More notably, these models rely heavily on semantic cues while often neglecting syntactic structures.

These studies demonstrate that while LLMs have greatly advanced code summarization under ideal conditions, they still struggle with obfuscated or poorly readable code. Motivated by this limitation, our work aims to enhance the robustness of code summarization models against obfuscation, addressing a critical gap in existing literature and improving real-world applicability.

\subsection{Robustness of Code Summarization}
Besides our work, there has been other research on improving the robustness of code summarization models on obfuscated code. For example, Jain et al.~\cite{jain2020contrastive} introduced ContraCode, a contrastive pre-training task for learning code representations. ContraCode pre-trains a neural network to distinguish functionally similar code snippets from non-equivalent distractors.
Similarly, Yan et al.~\cite{yan2021towards} proposed VECOS, a generative tree-based model specifically designed for code summarization. They hypothesize that previous models heavily rely on cross-attention between code tokens and summarized words, which can be problematic when meaningful identifiers are obscured. Their approach forces the representation of the eroded code to align with the representation of its original counterpart via variational inference. 
Jia et al.~\cite{jia2023clawsat} proposed CLAWSAT, which integrates contrastive learning with adversarial learning to co-optimize the robustness and accuracy of code language models. Their approach involves training the model on adversarial code to improve its robustness.

Although these approaches have demonstrated promising results, they often require the introduction of new pretraining tasks or modifications to the model architecture. In contrast, our approach follows the fine-tuning pipeline and utilizes the Transformer architecture commonly employed in LLMs, making only modest adjustments to the fine-tuning workflow.

\section{Conclusion}

In this paper, we proposed \approach, a readability-agnostic method for code summarization. \approach enhances the robustness of the fine-tuning process by marrying the ideas of meta-learning and curriculum learning.
It constructs datasets that have progressive difficulty in code comprehension and meta-updates the gradients on data with progressive difficulty, optimizing both the robustness and accuracy simultaneously. 
The experimental results show that \approach achieves a significant improvement in summarizing the poorly readable code compared to the baseline methods. Our work also provides insights into how to effectively fine-tune large language models.
In the future, we will explore more code obfuscations to inspect the effect of robust fine-tuning in a wider spectrum of curricula.

\bibliographystyle{IEEEtran}
\bibliography{custom}

\end{document}

%% file: tables/empirical_sota.tex
\begin{table*}[ht]
    \caption{Performance of state-of-the-art LLMs on code summarization for high and poor readability code. Results show BLEU and SBERT scores across two test datasets.}
    \centering
    \begin{tabular}{cc@{\hspace{2pt}}cc@{\hspace{2pt}}cc@{\hspace{2pt}}cc@{\hspace{2pt}}cc@{\hspace{2pt}}cc@{\hspace{2pt}}cc}
        \toprule
        \multirow{3}{*}{\bf Model} & \multicolumn{4}{c}{\bf MLRC} & \multicolumn{8}{c}{\bf CSN} \\
        & \multicolumn{2}{c}{\bf \cellcolor{gray!10} High} & \multicolumn{2}{c}{\bf \cellcolor{gray!50} Low} & \multicolumn{2}{c}{\bf \cellcolor{gray!10} Origin} & \multicolumn{2}{c}{\bf \cellcolor{gray!50} DCI} & \multicolumn{2}{c}{\bf \cellcolor{gray!50} FNE} & \multicolumn{2}{c}{\bf \cellcolor{gray!50} IRN} \\
        \cmidrule(lr){2-3} \cmidrule(lr){4-5} \cmidrule(lr){6-7} \cmidrule(lr){8-9} \cmidrule(lr){10-11} \cmidrule(lr){12-13}
        &  BLEU &  SBERT
        &  BLEU &  SBERT
        &  BLEU &  SBERT 
        &  BLEU &  SBERT 
        &  BLEU &  SBERT 
        &  BLEU &  SBERT \\
        \midrule
        \makecell[c]{Qwen-Max-0125} & 9.33 & 60.96 & 6.30 & 47.12 & 15.66 & 62.74 & 14.68 & 57.44 & 14.35 & 57.97 & 12.96 & 51.20 \\
        \makecell[c]{DeepSeek-V3-0324}   & 7.31 & 54.29 & 7.24 & 44.29 & 15.85 & 62.86 & 15.31 & 59.95 & 15.23 & 58.96 & 14.46 & 53.05 \\
        \makecell[c]{Gemini-2.0-Flash}   & 11.09 & 57.05 & 8.88 & 44.50 & 18.71 & 64.24 & 17.67 & 59.35 & 17.87 & 60.58 & 17.20 & 55.20 \\
        \makecell[c]{Claude-3.5-Sonnet-20241022}   & 8.43 & 54.66 & 8.77 & 46.78 & 14.40 & 62.08 & 13.97 & 57.28 & 13.60 & 58.83 & 12.79 & 53.24 \\
        \makecell[c]{GPT-4o-2024-11-20}     & 8.13 & 57.20 & 6.87 & 44.47 & 14.71 & 62.06 & 14.08 & 57.85 & 13.66 & 57.31 & 12.79 & 52.09 \\
        \bottomrule
    \end{tabular}
    \label{tab:sota}
\end{table*}

%% file: tables/empirical_prompt.tex
\begin{table*}[ht]
    \caption{Performance of DeepSeek-V3 with four prompt engineering methods on code summarization for high and poor readability code. Results show BLEU and SBERT scores across two test datasets.}
    \centering
    \begin{tabular}{cc@{\hspace{2pt}}cc@{\hspace{2pt}}cc@{\hspace{2pt}}cc@{\hspace{2pt}}cc@{\hspace{2pt}}cc@{\hspace{2pt}}cc}
        \toprule
        \multirow{3}{*}{\bf Method} & \multicolumn{4}{c}{\bf MLRC} & \multicolumn{8}{c}{\bf CSN} \\
        & \multicolumn{2}{c}{\bf \cellcolor{gray!10} High} & \multicolumn{2}{c}{\bf \cellcolor{gray!50} Low} & \multicolumn{2}{c}{\bf \cellcolor{gray!10} Origin} & \multicolumn{2}{c}{\bf \cellcolor{gray!50} DCI} & \multicolumn{2}{c}{\bf \cellcolor{gray!50} FNE} & \multicolumn{2}{c}{\bf \cellcolor{gray!50} IRN} \\
        \cmidrule(lr){2-3} \cmidrule(lr){4-5} \cmidrule(lr){6-7} \cmidrule(lr){8-9} \cmidrule(lr){10-11} \cmidrule(lr){12-13}
        &  BLEU &  SBERT 
        &  BLEU &  SBERT 
        &  BLEU &  SBERT 
        &  BLEU &  SBERT 
        &  BLEU &  SBERT 
        &  BLEU &  SBERT \\
        \midrule
        \makecell[c]{Zero-Shot} & 7.31 & 54.29 & 7.24 & 44.29 & 15.85 & 62.86 & 15.31 & 59.95 & 15.23 & 58.96 & 14.46 & 53.05 \\
        \makecell[c]{Few-Shot (3-shot)} & - & - & - & - & 13.11 & 60.82 & 13.71 & 59.39 & 12.78 & 57.32 & 12.61 & 53.42 \\
        \makecell[c]{Chain-of-Thought}  & 6.79 & 55.24 & 5.06 & 41.43 & 11.22 & 59.48 & 10.76 & 54.16 & 10.90 & 55.52 & 10.14 & 49.48 \\
        \makecell[c]{Critique}   & 7.13 & 54.80 & 7.22 & 38.64 & 9.28 & 56.67 & 9.23 & 49.70 & 8.89 & 53.62 & 7.56 & 46.42 \\
        \bottomrule
    \end{tabular}
    \label{tab:prompt}
\end{table*}

%% file: tables/empirical_variants.tex
\begin{table*}[ht]
    \caption{Performance of Qwen2.5 series with three model variants on code summarization for high and poor readability code. Results show BLEU and SBERT scores across two test datasets.}
    \centering
    \begin{tabular}{cc@{\hspace{2pt}}cc@{\hspace{2pt}}cc@{\hspace{2pt}}cc@{\hspace{2pt}}cc@{\hspace{2pt}}cc@{\hspace{2pt}}cc}
        \toprule
        \multirow{3}{*}{\bf Model} & \multicolumn{4}{c}{\bf MLRC} & \multicolumn{8}{c}{\bf CSN} \\
        & \multicolumn{2}{c}{\bf \cellcolor{gray!10} High} & \multicolumn{2}{c}{\bf \cellcolor{gray!50} Low} & \multicolumn{2}{c}{\bf \cellcolor{gray!10} Origin} & \multicolumn{2}{c}{\bf \cellcolor{gray!50} DCI} & \multicolumn{2}{c}{\bf \cellcolor{gray!50} FNE} & \multicolumn{2}{c}{\bf \cellcolor{gray!50} IRN} \\
        \cmidrule(lr){2-3} \cmidrule(lr){4-5} \cmidrule(lr){6-7} \cmidrule(lr){8-9} \cmidrule(lr){10-11} \cmidrule(lr){12-13}
        &  BLEU &  SBERT 
        &  BLEU &  SBERT 
        &  BLEU &  SBERT 
        &  BLEU &  SBERT 
        &  BLEU &  SBERT 
        &  BLEU &  SBERT \\
        \midrule
        \makecell[c]{Qwen2.5-32B} & 4.07 & 38.41 & 3.04 & 23.24 & 5.40 & 26.49 & 5.74 & 20.24 & 4.96 & 21.80 & 4.02 & 13.69 \\
        \makecell[c]{Qwen2.5-32B-Instruct} & 10.22 & 60.35 & 6.44 & 42.14 & 14.29 & 61.68 & 12.91 & 54.84 & 13.04 & 56.78 & 11.92 & 49.65 \\
        \makecell[c]{QwQ-32B}  & 5.75 & 48.97 & 5.93 & 44.19 & 11.98 & 60.78 & 12.21 & 57.80 & 11.23 & 56.70 & 12.34 & 51.42 \\
        \bottomrule
    \end{tabular}
    \label{tab:variants}
\end{table*}

%% file: tables/result_fne_irn_dscoder.tex
\begin{table*}[ht]
    \caption{Results of various methods in the semantic-obfuscation curriculum when backend by \textbf{DeepSeek-Coder}. ``Origin", ``FNE", and ``IRN" denote the test sets of three difficulty levels respectively. 
    The column ``AVG'' shows the average BLEU and SBERT scores across three test sets, with parenthetical values denoting the corresponding improvements over direct fine-tuning on the original dataset. The improvements of \approach over Fine-tuning are statistically significant across all results (Wilcoxon signed-rank test, $p < 0.05$).
    \\ $^*$For zero-shot summarization, we use the default \emph{fill-in-the-middle} methods in the models' guideline documentation$^{1,2}$}
    \centering
    \resizebox{0.75\textwidth}{!}{
    \begin{tabular}{cc@{\hspace{2pt}}cc@{\hspace{2pt}}cc@{\hspace{2pt}}cc@{\hspace{2pt}}cc@{\hspace{2pt}}cc@{\hspace{2pt}}}
        \toprule
        \bf Training & \multicolumn{2}{c}{\cellcolor{gray!10} \bf Origin} & \multicolumn{2}{c}{\cellcolor{gray!20} \bf FNE} & \multicolumn{2}{c}{\cellcolor{gray!30} \bf IRN} & \multicolumn{2}{c}{\bf AVG} \\
        \bf Method & \bf BLEU & \bf SBERT & \bf BLEU & \bf SBERT & \bf BLEU & \bf SBERT & \bf BLEU & \bf SBERT \\
        \midrule
        Zero-Shot      & 18.81 & 56.56 & 13.11 & 42.44 & 7.01 & 23.13 & 12.98 & 40.71 \\
        \hline
        FT\textsubscript{Origin} & 23.87 & 61.18 & 20.93 & 53.54 & 15.78 & 37.70 & 20.19 & 50.81 \\
        FT\textsubscript{All} & 24.39 & 61.34 & 22.28 & 55.88 & 21.22 & 51.48 & 22.63 (+2.44) & 56.23 (+5.42) \\
        \hline
        CL             & 23.77 & 60.97 & 21.87 & 55.59 & 20.72 & 51.08 & 22.12 (+1.93) & 55.88 (+5.07) \\
        CLAWSAT        & 23.46 & 60.94 & 21.58 & 55.37 & 20.54 & 50.81 & 21.86 (+1.67) & 55.71 (+4.90) \\
        \hline
        \approach & \textbf{24.94} & \textbf{61.87} & \textbf{23.18} & \textbf{56.95} & \textbf{22.39} & \textbf{53.12} & \textbf{23.50} (\textbf{+3.31}) & \textbf{57.31} (\textbf{+6.50}) \\
        \bottomrule
    \end{tabular}
    }
    \label{tab:result:dscoder_fne_irn_curriculum}
\end{table*}

%% file: tables/result_dci_dscoder.tex
\begin{table*}[ht]
    \caption{Results of various methods in the semantic-interference curriculum backend by \textbf{DeepSeek-Coder}. ``Origin", ``DCI5", and ``DCI10" denote the test sets of three difficulty levels respectively. The improvements of \approach over Fine-tuning are statistically significant across all results (Wilcoxon signed-rank test, $p < 0.05$).}
    \centering
    \resizebox{0.75\textwidth}{!}{
    \begin{tabular}{cc@{\hspace{1pt}}cc@{\hspace{1pt}}cc@{\hspace{1pt}}cc@{\hspace{2pt}}cc@{\hspace{2pt}}cc@{\hspace{2pt}}}
        \toprule
        \bf Training & \multicolumn{2}{c}{\cellcolor{gray!10} \bf Origin} & \multicolumn{2}{c}{\cellcolor{gray!20} \bf DCI5} & \multicolumn{2}{c}{\cellcolor{gray!30} \bf DCI10} & \multicolumn{2}{c}{\bf AVG} \\
        \bf Method & \bf BLEU & \bf SBERT & \bf BLEU & \bf SBERT & \bf BLEU & \bf SBERT & \bf BLEU & \bf SBERT \\
        \midrule
        Zero-Shot      & 18.81 & 56.56 & 18.41 & 55.56 & 18.03 & 54.59 & 18.42 & 55.57 \\
        \hline
        FT\textsubscript{Origin} & 23.87 & 61.18 & 23.39 & 60.10 & 23.08 & 59.07 & 23.45 & 60.12 \\
        FT\textsubscript{All}     & 24.39 & 61.34 & 24.07 & 60.44 & 23.86 & 59.83 & 24.11 (+0.66) & 60.54 (+0.42) \\
        \hline
        CL             & 23.96 & 61.29 & 23.73 & 60.36 & 23.51 & 59.74 & 23.73 (+0.28) & 60.46 (+0.34) \\
        CLAWSAT        & 23.71 & 61.18 & 23.51 & 60.34 & 23.26 & 59.67 & 23.49 (+0.04) & 60.40 (+0.28) \\
        \hline
        \approach & \textbf{24.91} & \bf 61.94 & \textbf{24.75} & \bf 61.23 & \textbf{24.65} & \textbf{60.71} & \textbf{24.77} (\textbf{+1.32}) & \textbf{61.29} (\textbf{+1.17}) \\
        \bottomrule
    \end{tabular}
    }
    \label{tab:result:dscoder_dci_curriculum}
\end{table*}

%% file: tables/result_fne_irn_qwen.tex
\begin{table*}[ht]
    \centering
    \caption{Results of various methods in the semantic-obfuscation curriculum when backend by \textbf{Qwen2.5-Coder}. The improvements of \approach over Fine-tuning are statistically significant across all results (Wilcoxon signed-rank test, $p < 0.05$).}
    \resizebox{0.75\textwidth}{!}{
    \begin{tabular}{cc@{\hspace{2pt}}cc@{\hspace{2pt}}cc@{\hspace{2pt}}cc@{\hspace{2pt}}cc@{\hspace{2pt}}cc@{\hspace{2pt}}}
        \toprule
        \bf Training & \multicolumn{2}{c}{\cellcolor{gray!10} \bf Origin} & \multicolumn{2}{c}{\cellcolor{gray!20} \bf FNE} & \multicolumn{2}{c}{\cellcolor{gray!30} \bf IRN} & \multicolumn{2}{c}{\bf AVG} \\
        \bf Method & \bf BLEU & \bf SBERT & \bf BLEU & \bf SBERT & \bf BLEU & \bf SBERT & \bf BLEU & \bf SBERT \\
        \midrule
        Zero-Shot      & 17.60 & 56.57 & 13.48 & 46.74 & 8.42 & 30.56 & 13.17 & 44.62 \\
        \hline
        FT\textsubscript{Origin} & 22.75 & 60.90 & 20.45 & 54.97 & 17.88 & 46.59 & 20.36 & 54.15 \\
        FT\textsubscript{All}  & 23.28 & 61.08 & 21.38 & 55.94 & 20.57 & 52.07 & 21.74(+1.38) & 56.36(+2.21) \\
        \hline
        CL             & 22.73 & 60.86 & 21.16 & 55.97 & 20.16 & 51.91 & 21.35(+0.99) & 56.25(+2.10) \\
        CLAWSAT        & 22.59 & 60.45 & 20.85 & 55.34 & 19.89 & 50.98 & 21.11(+0.75) & 55.59(+1.44) \\
        \hline
        \approach  & \textbf{23.60} & \textbf{61.31} & \textbf{21.98} & \textbf{56.78} & \textbf{21.27} & \textbf{53.20} & \textbf{22.28}(\textbf{+1.92}) & \textbf{57.10}(\textbf{+2.95}) \\
        \bottomrule
    \end{tabular}
    }
    \label{tab:result:qwencoder_fne_irn_curriculum}
\end{table*}

%% file: tables/result_dci_qwen.tex
\begin{table*}[ht]
    \centering
    \caption{Results of various methods in the semantic-interference curriculum backend by \textbf{Qwen2.5-Coder}. The improvements of \approach over Fine-tuning are statistically significant across all results (Wilcoxon signed-rank test, $p < 0.05$).}
    \resizebox{0.75\textwidth}{!}{
    \begin{tabular}{cc@{\hspace{2pt}}cc@{\hspace{2pt}}cc@{\hspace{2pt}}cc@{\hspace{2pt}}cc@{\hspace{2pt}}}
        \toprule
        \bf Training & \multicolumn{2}{c}{\cellcolor{gray!10} \bf Origin} & \multicolumn{2}{c}{\cellcolor{gray!20} \bf DCI5} & \multicolumn{2}{c}{\cellcolor{gray!30} \bf DCI10} & \multicolumn{2}{c}{\bf AVG}  \\
        \bf Method & \bf BLEU & \bf SBERT & \bf BLEU & \bf SBERT & \bf BLEU & \bf SBERT & \bf BLEU & \bf SBERT \\
        \midrule
        Zero-Shot      & 17.60 & 56.57 & 16.90 & 54.81 & 16.53 & 53.57 & 17.01 & 54.98  \\
        \hline
        FT\textsubscript{Origin} & 22.75 & 60.90 & 22.40 & 59.60 & 22.10 & 58.77 & 22.42 & 59.76 \\
        FT\textsubscript{All}     & 23.28 & 61.08 & 23.15 & 60.27 & 23.13 & 59.58 & 23.19 (+0.77) & 60.31 (+0.55) \\
        \hline
        CL             & 22.92 & 61.02 & 22.71 & 60.35 & 22.69 & 59.55 & 22.77 (+0.35) & 60.31 (+0.55) \\
        CLAWSAT        & 22.84 & 60.93 & 22.77 & 60.20 & 22.78 & 59.54 & 22.80 (+0.38) & 60.22 (+0.46) \\
        \hline
        \approach  & \textbf{23.78} & \textbf{61.49} & \textbf{23.78} & \textbf{61.08} & \textbf{24.11} & \textbf{60.54} & \textbf{23.89}(\textbf{+1.47}) & \textbf{61.04} (\textbf{+1.28}) \\
        \bottomrule
    \end{tabular}
    }
    \label{tab:result:qwencoder_dci_curriculum}
\end{table*}